\documentclass[letterpaper]{article} 
\usepackage[]{aaai25}  
\usepackage{times}  
\usepackage{helvet}  
\usepackage{courier}  
\usepackage{multirow}  
\usepackage{array}     
\usepackage{amsmath}   
\usepackage[hyphens]{url}  
\usepackage{graphicx} 
\usepackage{natbib}  
\usepackage{caption} 
\usepackage{algorithm}
\usepackage{algorithmic}

\usepackage{newfloat}
\usepackage{listings}
\DeclareCaptionStyle{ruled}{labelfont=normalfont,labelsep=colon,strut=off} 
\floatstyle{ruled}
\newfloat{listing}{tb}{lst}{}
\floatname{listing}{Listing}
\title{When Life Paths Cross: Extracting Human Interactions in Time and Space from Wikipedia}
\title{When Life Paths Cross: Extracting Human Interactions in Time and Space from Wikipedia}
\author {
    Zhongyang Liu\textsuperscript{\rm 1},
    Ying Zhang\textsuperscript{\rm 1,2},
    Xiangyi Xiao\textsuperscript{\rm 1},
    Wenting Liu\textsuperscript{\rm 1},
    Yuanting Zha\textsuperscript{\rm 1},
    Haipeng Zhang\textsuperscript{\rm 1}
}
\affiliations {
    \textsuperscript{\rm 1}ShanghaiTech University  
    \quad \textsuperscript{\rm 2}Ant Group 
    \\  
    \{liuzy12024,zhangying12022,xiaoxy,liuwt,zhayt2022,zhanghp\}@shanghaitech.edu.cn

}

\usepackage{bibentry}

\usepackage{booktabs}
\usepackage{color}

\usepackage{amssymb}
\usepackage{amsmath}

\begin{document}

\maketitle

\begin{abstract}
Interactions among notable individuals—whether examined individually, in groups, or as networks—often convey significant messages across cultural, economic, political, scientific, and historical perspectives. By analyzing the times and locations of these interactions, we can observe how dynamics unfold across regions over time. However, relevant studies are often constrained by data scarcity, particularly concerning the availability of specific location and time information. To address this issue, we mine millions of biography pages from Wikipedia, extracting 685,966 interaction records in the form of \textit{(Person1, Person2, Time, Location)} interaction quadruplets. The key elements of these interactions are often scattered throughout the heterogeneous crowd-sourced text and may be loosely or indirectly associated. We overcome this challenge by designing a model that integrates attention mechanisms, multi-task learning, and feature transfer methods, achieving an F1 score of \textbf{86.51\%}, which outperforms baseline models. We further conduct an empirical analysis of intra- and inter-party interactions among political figures to examine political polarization in the US, showcasing the potential of the extracted data from a perspective that may not be possible without this data. We make our code, the extracted interaction data, and the \textit{WikiInteraction} dataset of 4,507 labeled interaction quadruplets publicly available\footnote{https://anonymous.4open.science/r/FALCON-7EF9. ~For now, we release a subset of the extracted interaction data. The remaining data and the full \textit{WikiInteraction} dataset will be made available upon acceptance.}.


\end{abstract}

%

\section{Introduction}
Interpersonal interactions, especially among notable individuals, reveal insights into the cultural, economic, political, scientific, and historical perspectives of human society~\cite{jackson2011overview,o2014opened, cruz2017politician,fuller2021structuring}, as seen in the interactions of scientists~\cite{newman2001scientific,fortunato2018science}, politicians~\cite{hsu2012mapping,plotkowiak2013german}, and authors~\cite{borner2004simultaneous,sun2011co}. 
While smaller-scale datasets of real-world interactions exist~\cite{illenberger2013role,kossinets2006empirical}, and online behavioral data has historically been relatively easy to obtain from social media platforms such as Facebook and Twitter (now X) to support large-scale social and information network analysis~\cite{kleinberg2013analysis}, there remains a significant lack of large-scale, real-world interaction datasets. 
Although some studies attempt to address this issue by employing heuristic methods or neural network techniques to automate the extraction of individual interaction information from text~\cite{tang2007social,gergaud2016brief,tao2020character}, they overlook time and location—two critical attributes of interactions that influence how social dynamics unfold over time \cite{barabasi2002evolution,kossinets2006empirical}, across regions \cite{onnela2011geographic,crandall2010inferring}. If time and location data were available, spatial-temporal graph neural network techniques, commonly used in urban modeling~\cite{jin2023spatio}, could be employed to gain a deeper understanding of human dynamics, extending beyond online behaviors.

\begin{figure}[!tbp]
    \centering
    \includegraphics[width=0.95\linewidth]{{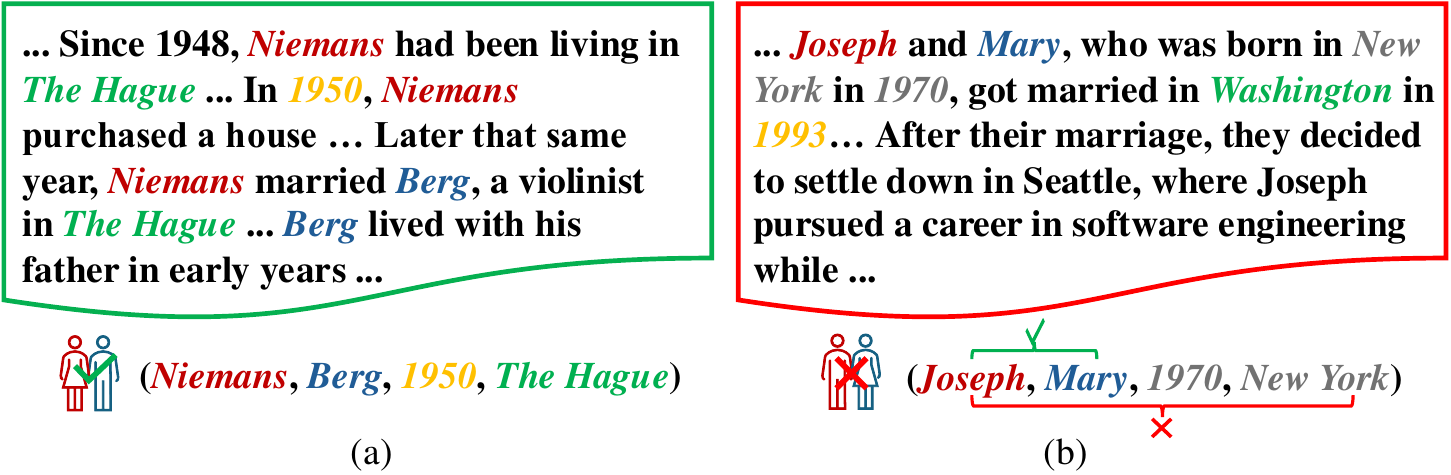}}
    
    \caption{Example of extracted quadruples and their contexts. (a) An example of correct spatio-temporal interaction.
(b) An example of incorrect spatio-temporal interaction.}
    \label{fig:sample}
    \vspace{-16pt}
\end{figure}

The task becomes extracting the correct \textit{(Person1, Person2, Time, Location)} interaction quadruplets from a text corpus, where \textit{Person1} and \textit{Person2} interact at \textit{Time} in \textit{Location}. 
According to prior research~\cite{8529125,labatut2019extraction}, we define an interaction as a direct action between two individuals (e.g., marriage, collaboration, competition, conversation).
We choose Wikipedia's biography pages as the extraction source, which contains abundant spatio-temporal information related to human life~\cite{suchanek2008yago}. For example, as shown in Figure~\ref{fig:sample} (a), the quadruplet (\textit{Niemans}, \textit{Berg}, \textit{1950}, \textit{The Hague}) represents an interaction between \textit{Niemans} and \textit{Berg} that occurred in \textit{The Hague} in \textit{1950}.

Extracting these quadruplets is a nontrivial task. Building on previous studies~\cite{tang2007social,gergaud2016brief,tao2020character} that extract social networks by classifying text to determine interactions between individuals (\textit{Person1} and \textit{Person2}), a heuristic approach would employ a Named Entity Recognition (NER) tool to detect temporal and locational entities near them. However, this method often fails because \textit{Person1} and \textit{Person2} may not even have had an interaction in the first place. Even if they did, the relevant \textit{Time} and \textit{Location} may not be the closest entities identified. Another study~\cite{zhang2025paths} generates interaction quadruples by combining spatio-temporal co-occurring triples of trajectories (person, time, location). However, co-occurrence does not equal real interaction, which makes this method difficult to capture real interaction relationships and leads to the accumulation of model errors. The core challenge lies in how to precisely associate individuals with the specific spatiotemporal entities corresponding to their real interactions. For example, as shown in Figure~\ref{fig:sample} (b), the heuristic method identifies an interaction (\textit{Joseph}, \textit{Mary}, \textit{1970}, \textit{New York}), but closer examination reveals they actually interacted in \textit{Washington} in \textit{1993}.

Moreover, these entities are often scattered across different sections of the text, complicating their associations. In Figure~\ref{fig:sample}(a), \textit{Niemans} is first located in \textit{The Hague}, then referenced in \textit{1950}, and finally mentioned in an interaction with \textit{Berg}. Thus, considering the context of these mentions is crucial for accuracy, and not every occurrence holds equal significance; for instance, a second mention of \textit{Berg} does not contribute to our assessment.

The task is transformed into identifying the correct spatio-temporal information quadruplets from candidate quadruplets generated by combining potential NER results from the original corpus.
To simplify, we split the problem into two related tasks: (1) a main task that determines if an interaction occurs between \textit{Person1} and \textit{Person2} given a \textit{Time} and \textit{Location}, and (2) an auxiliary task that verifies their geo-temporal co-occurrence at the same \textit{Location} and \textit{Time}.
The auxiliary task can provide reliable spatio-temporal evidence, supporting the main task.
This naturally fits a multi-task learning framework, where the joint training process forces the model to capture spatio-temporal correlations that are crucial for both tasks.
Additionally, transfer learning can enhance performance by incorporating features learned from the auxiliary task into interaction detection. We show in the experiments that a synergistic effect between multi-task and transfer learning improves the overall process.

As noted, focusing on relationships between multiple entities scattered throughout the text is crucial, making the contextual information of each entity particularly important. Recent studies, like R-Bert~\cite{wu2019enriching}, effectively capture this contextual information. However, R-Bert struggles with the varying importance of the same entity appearing multiple times in different positions. To address this, we introduce an attention mechanism based on R-Bert to aggregate contextual information across positions, dynamically adjusting weights for better integration of semantic information. We call this enhanced approach AR-Bert (Attention-enhanced R-Bert), which serves as our feature extractor.
 
In this paper, we propose \textbf{FALCON} (AR-Bert model utilizing \textbf{F}eature Tr\textbf{A}nsfer and Multi-Task \textbf{L}earning strategies for extracting spatio-temporal Life Intera\textbf{C}tio\textbf{ON}s) for extracting spatio-temporal interactions. Initially, we use a heuristic method to extract quadruples (\textit{Person1}, \textit{Person2}, \textit{Time}, \textit{Location}) as classification candidates, where a correct quadruple indicates interaction at the specified \textit{Time} and \textit{Location}. These quadruples and their contexts are input to our model, which classifies them as ``interaction'' or ``not interaction.'' The model is therefore evaluated under binary classification metrics. We annotate a new spatio-temporal interaction dataset \textit{WikiInteraction} with 4,507 quadruples (7:1:2 split for train, validation, and test). Each quadruple is decomposed into two presence triplets for annotation, resulting in 9,014 triplets (\textit{Person}, \textit{Time}, \textit{Location}), which can be viewed as life trajectories~\cite{zhang2025paths}. We define the auxiliary task as a trajectory task, categorizing these triplets into ``trajectory'' and ``not trajectory'' for multi-task learning and feature transfer.

Additionally, we apply the model to the entire English Wikipedia. Based on the extracted data, we showcase how post-processing can be used to determine the types of interactions for specific analysis scenarios, such as political interactions, thereby enhancing its usability. We conduct an empirical analysis of political polarization in the US, focusing on intra- and inter-party interactions among political figures to demonstrate the potential of our data.

We summarize our contributions as follows:
\begin{itemize}
     \item We formally introduce the task of extracting spatio-temporal interactions from Wikipedia biographies and construct a curated dataset \textit{WikiInteraction} for this task. While our experiments focus on Wikipedia biographies, the proposed methods can also be applied to other textual materials.

    \item We design an effective framework, \textbf{FALCON}, which combines the ideas of multi-task learning, transfer learning, and using our improved AR-Bert as a feature extractor. \textbf{FALCON} achieves an F1 score of 86.51\% on the dataset, outperforming all baselines and generalizes well on another important source of biographies, Encyclopedia Britannica.

    \item   We extract 685,966  interactions, which constitutes the largest existing spatio-temporal interaction dataset. 
    Additionally, we conduct an empirical analysis of political polarization in the US, focusing on intra- and inter-party interactions to demonstrate the potential of our data.
    Our code, the annotated dataset, and the extracted Wikipedia interactions are publicly available.

\end{itemize}

\section{Related Work}

\subsection{Analysis of Interaction Data}

Interaction data has significant application value in social sciences: it can not only reveal deep social culture, economy, politics, and running mechanisms~\cite{jackson2008social,o2014opened, cruz2017politician,fuller2021structuring}, but also parse the behavior patterns of specific clusters such as research groups and political groups~\cite{newman2001scientific,fortunato2018science, hsu2012mapping, plotkowiak2013german}. Empirical analysis based on interaction data can better promote the resolution of real-world problems. For instance, \citet{jeong2024empirical} has improved rural medical services accordingly, and \citep{li2024impact} has facilitated the formulation of community revitalization strategies.

However, the existing interaction data generally have the drawbacks of limited time and insufficient spatial coverage. In contrast, spatio-temporal interaction data can once again create novel perspectives by providing multi-dimensional information increments to overcome the traditional limitations. Typical cases include: \citet{barabasi2002evolution} tracks the dynamic evolution of scientific research collaboration networks along with the development of disciplines. \citet{onnela2011geographic} demonstrates the shaping effect of geographical location on the structure of social networks. \citet{jin2023spatio} utilizes spatio-temporal graph models to deepen human behavior cognition.

\subsection{Extraction of Interaction Data}



Extracting interaction data between individuals from text has been a key research focus. Early methods used rule-based approaches to identify individuals~\cite{Backstrom2006GroupFI, tang2007social}, followed by NER-based techniques that identify character entities and extract interactions through co-occurrence rules or trigger words~\cite{gergaud2016brief, 9037669, agarwal2016social}. Recent advances have employed deep learning methods (CNN and Bi-LSTM) to improve extraction performance~\cite{8529125,tao2020character}.

However, existing methods primarily focus on detecting whether interactions occur, neglecting temporal and spatial dimensions. To address this limitation, we propose a novel task for extracting spatio-temporal interaction information and design a multi-task learning model to solve it.

\section{Formulation of Task and Annotation}
We define the task as determining whether two individuals interact at a specified time and location within a given text segment. For each candidate quadruple $(Person1, Person2, Time, Location)$ extracted from a paragraph, a model $f$ classifies it as $y=1$ (interaction exists) or $y=0$ (no interaction), where ``interaction" requires co-occurrence with a meaningful connection (e.g., conversation, joint activity).

To build the dataset, we annotated 4,507 candidate quadruples using a three-person team (two annotators and one checker). Each quadruple was split into two trajectory triples $(Person, Time, Location)$, resulting in 9,014 triples for auxiliary trajectory labeling. Positive labels indicate valid interactions/trajectories; negative labels indicate invalid cases. The label distribution is shown in Table~\ref{tab:Data_situation}. We have provided a detailed introduction to the acquisition process of the dataset (such as the acquisition of candidate quadruples) in the Appendix.

\begin{table}[!htb]
  \centering
    \begin{tabular}{lllll}\toprule
     Type&Positive  &Negative  &Total \\\midrule
    Interaction &2,351 &2,156 &4,507 \\
    Trajectory &5,730 &3,284 &9,014 \\
\bottomrule
    \end{tabular}%
    \vspace{-5pt}
    \caption{Distribution of the $WikiInteraction$ Dataset.}
    \label{tab:Data_situation}%
    \vspace{-15pt}
\end{table}%



\section{Method}

\begin{figure*}[!h]
    \centering
    \includegraphics[width=0.85\linewidth]{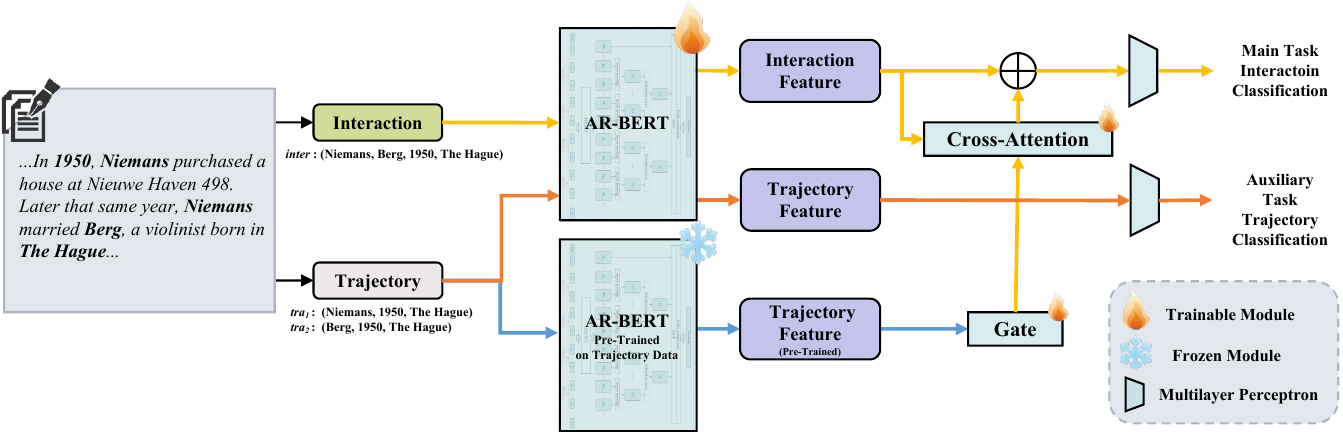}
    \caption{The framework of our method. }
    \label{fig:The framework of our method.}
    \vspace{-10pt} 
\end{figure*}

Our FALCON framework (Fig.~\ref{fig:The framework of our method.}) processes candidate interactions comprising a text segment $s$ and a quadruple $q$. First, $q$ is decomposed into two trajectory triplets ($t_1$, $t_2$). The quadruple $q$ drives the main \textbf{Interaction Classification} task, while the triplets $t_1$ and $t_2$ drive the auxiliary \textbf{Trajectory Classification} task.

We employ a trainable AR-BERT model as the primary feature extractor for both tasks, yielding interaction features $H^{\prime}_{inter}$ and trajectory features $H^{\prime}_{tra}$. Simultaneously, a \textit{frozen} AR-BERT model, pre-trained solely on trajectory data, extracts trajectory features $H_{\text{tra}}$. Transfer learning is incorporated by fusing $H^{\prime}_{inter}$ and $H_{\text{tra}}$ into $H^{\prime}_{fusion}$ for the final interaction prediction. The model is jointly trained using multi-task learning.

The following sections detail: (1) the \textbf{AR-BERT} feature extractor, (2) the main \textbf{Interaction Classification} task, (3) the auxiliary \textbf{Trajectory Classification} task, and (4) the \textbf{Multi-Task Learning} strategy.

\subsection{AR-BERT}
\begin{figure}[htbp]
    \centering
    \includegraphics[width=0.85\linewidth]{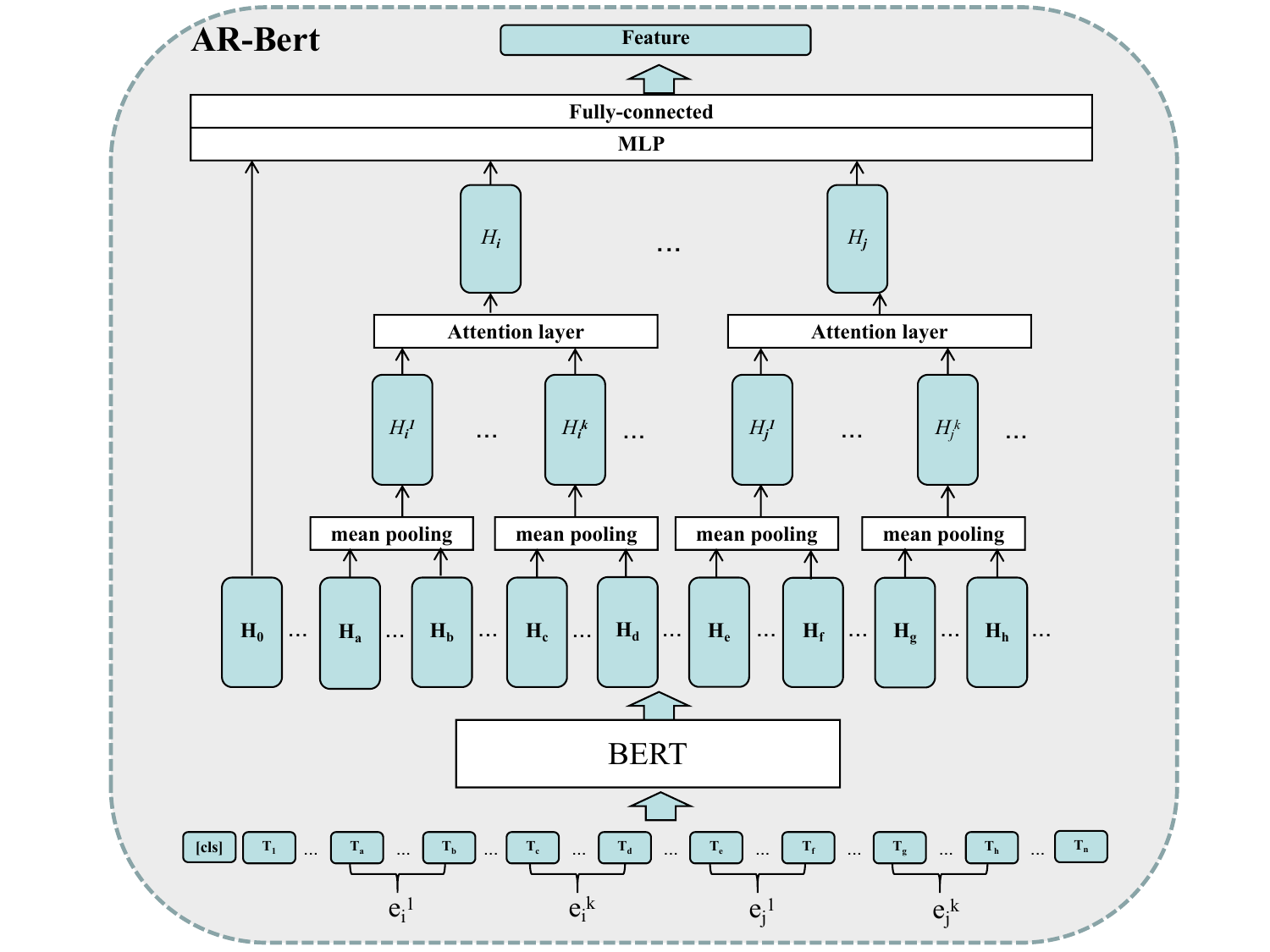}
    \caption{The architecture of AR-BERT.}
    \label{fig:AR-BERT}
    \vspace{-10pt} 
\end{figure}

The architecture of AR-BERT, designed to enhance BERT's understanding of entities in a given text, is illustrated in Figure~\ref{fig:AR-BERT}.
\subsubsection{Input and Embedding Representation}
Given an input text segment $s$ and a set of entities $E = \{e_1, e_2, \ldots, e_n\}$, we first obtain the embedding representation using BERT. The final hidden state of the \texttt{[CLS]} token is used as the sentence's overall representation.

\subsubsection{Special Marker Insertion}
We enhance BERT's ability to capture entity information by inserting special tokens around each entity. For example, for entities such as \textit{Person1}, \textit{Person2}, \textit{Time} and \textit{Location}, we insert the following markers: \textbf{Person1}:  `\#'; \textbf{Person2}: `\$' ; \textbf{Time}: `*' ; \textbf{Location}: `\&'. The sentence is then transformed with these markers.

\subsubsection{Entity Information Representation}

Each entity's representation is obtained by mean pooling the hidden states corresponding to each occurrence of the entity. The pooled vector for the $k$-th occurrence of entity $e_i$ is given by:
$ 
H_i^k = \frac{1}{d - c + 1} \sum_{t=c}^{d} \mathbf{H}_t,
 $
where $c$ and $d$ represent the start and end positions of the entity in the final hidden state output from BERT, respectively. Next, the attention mechanism is applied to fuse the representations:
$ 
H_i = \text{attn} \left( H_i^1, H_i^2, \ldots, H_i^v \right).
 $
The importance of each position is computed as:
$ 
w_i^k = \tanh \left( \boldsymbol{W}_{attn} H_i^k + \boldsymbol{b}_{attn} \right).
 $
The importance score of each position is then calculated as:
$ 
\delta_i^k = \frac{w_i^k}{\sum_{u=1}^{v} w_i^u}.
 $
The final entity embedding is:
$ 
H_i = \sum_{k=1}^{v} \delta_i^k H_i^k.
 $

\subsubsection{Fully Connected Layer Processing}

Each entity representation is passed through an activation function and fully connected layer:$ 
H_{e_i}^{\prime} = \mathbf{W}_i \left( \tanh(H_{e_i}) \right) + \mathbf{b}_i.
 $
Similarly, for the \texttt{[CLS]} token, we compute:
$ 
H_0^{\prime} = \mathbf{W}_0 \left( \tanh(\mathbf{H}_0) \right) + \mathbf{b}_0.
 $

\subsubsection{Feature Concatenation}
Finally, the embeddings of the \texttt{[CLS]} token and all entities are concatenated to form the final feature vector:
$ 
H^{\prime} = \text{concat}\left(H_0^{\prime}, H_{e_1}^{\prime}, H_{e_2}^{\prime}, \ldots, H_{e_n}^{\prime}\right).
 $
The feature vector is then used as input for subsequent tasks.

\subsubsection{AR-BERT Function}
The entire AR-BERT process can be summarized as:
$ 
H^{\prime} = \mathbf{AR\_BERT}(s, E),
 $
where $H^{\prime} \in \mathbb{R}^{(n+1)d}$.

\subsection{Main Task: Interaction Classification}
\subsubsection{Input}
Input includes text segment $s_{inter}$ and quadruple $E_q=(\textit{Person1}, \textit{Person2}, \textit{Time}, \textit{Location})$.

\subsubsection{Feature Extraction}
Interaction feature $H^{\prime}_{inter}$ is derived via AR-BERT.
\begin{equation}
H^{\prime}_{inter} = \mathbf{AR\_BERT}(s_{inter}, E_q),
\label{eq:AR-BERT_Function_Simple_Modified}
\end{equation}
with $H^{\prime}_{inter} \in \mathbb{R}^{5d}$.
Trajectory accuracy is vital for interaction correctness. We integrate trajectory feature $H^{\prime}_{tra}$ via feature transfer, yielding fused feature $H^{\prime}_{fusion}$.
A trajectory extractor $f_{\text{tra}}$ is trained on a separate trajectory dataset~\cite{zhang2025paths}, defined as:  
$ 
f_{\text{tra}}(s, E) = \mathbf{MLP}(\mathbf{AR\_BERT}(s, E)).
$
$f_{\text{tra}}$ outputs $\mathbb{R}^d$ features and does not participate in backpropagation.
For interaction $\textit{inter}=(s_{inter}, E_q)$, it splits into two trajectories: $\textit{tra}_1 =(s_{inter},E_{t_1})$ and $\textit{tra}_2 =(s_{inter},E_{t_2})$, with $E_{t1}=(\textit{Person1}, \textit{Time}, \textit{Location})$ and $E_{t2}=(\textit{Person2}, \textit{Time}, \textit{Location})$. Trajectory features $H_{\text{tra}_i}^{t}$ are extracted via $f_{\text{tra}}$.
\begin{equation}
H_{\text{tra}_i}^{t} = f_{\text{tra}}\left(tra_i\right). 
\label{eq:featureConcatenation}
\end{equation}
To incorporate trajectory features, we use gating and cross-attention.
Gating computes weights via sigmoid and matrix $\mathbf{W}_{\text{gate}}$, then filters features.
\begin{align}
\mathbf{gate}_{\text{tra}_i} &= \sigma\left(\mathbf{W}_{\text{gate}} H_{\text{tra}_i}^{t}\right),  
\label{eq:featureConcatenation}
\end{align} 
\begin{equation}
H_{\text{tra}_i}^g \&= \mathbf{gate}_{\text{tra}_i} \odot H_{\text{tra}_i}^{t}, 
\label{eq:featureConcatenation}
\end{equation}
where $i=1,2 $ and $\mathbf{W}_{\text{gate}}   \in \mathbb{R}^{d \times d}$.
Cross-attention applies query matrix $\mathbf{W}_Q$ to $H_{\text{inter}}^{\prime}$, with $H_{\text{tra}_i}^g$ as key and value.
\begin{equation}
Q = \mathbf{W}_Q\left(H_{\text{inter}}^{\prime}\right), 
\label{eq:featureConcatenation}
\end{equation}
\begin{equation}
H_{\text{tra}_i}^a \&= \text{Softmax}\left(\frac{QH_{\text{tra}_i}^g}{d}\right)H_{\text{tra}_i}^g,
\label{eq:featureConcatenation}
\end{equation}
where $i=1,2 $ and $\mathbf{W}_Q    \in \mathbb{R}^{d \times 5d}$.
Fused feature $H_{fusion}^{\prime}$ is obtained by concatenation.
\begin{equation}
H_{fusion}^{\prime} = \text{concat}\left(H_{\text{inter}}^{\prime} ,H_{\text{tra}_1}^a,  H_{\text{tra}_2}^a\right),
\label{eq:featureConcatenation}
\end{equation}

\subsubsection{Classification Head and Loss}
Label $\hat{y}_{inter}$ is derived via linear layer and softmax.
\begin{equation}
\hat{y}_{inter}=\text{Softmax} \left(W_{inter}H_{\text{fusion}}^{\prime}\right),
\end{equation}
where $W_{inter} \in \mathbb{R}^{2 \times 7d}$.
Loss $\mathcal{L}_{\text{inter}}$ is computed as:
\begin{align}
\mathcal{L}_{\text {inter }}&=-\sum_j^J y_{inter}^j \log \left(\hat{y}_{inter}^j\right)\nonumber  \\ 
&+\left(1-y_{inter}^j\right) \log \left(1-\hat{y}_{inter}^j\right).
\label{eq:featureConcatenation}
\end{align}

\subsection{Auxiliary Task: Trajectory Classification}
\subsubsection{Input}
Interaction data splits into $\text{tra}_1 = (s_{inter}, E_{t_1})$ and $\text{tra}_2 = (s_{inter}, E_{t_2})$, with $E_{t_1}=(\textit{Person1}, \textit{Time}, \textit{Location})$ and $E_{t_2}=(\textit{Person2}, \textit{Time}, \textit{Location})$.

\subsubsection{Feature Extraction}
Trajectory features $H^{\prime}_{tra_i}$ via AR-BERT:
\begin{equation}
H^{\prime}_{tra_i} = \mathbf{AR\_BERT}(t_{tra_i}, E_{t_i}), 
\label{eq:AR-BERT_Function_Simple_Modified}
\end{equation}
where $i \in \{1,2\}, H^{\prime}_{tra_i} \in \mathbb{R}^{4d}$.

\subsubsection{Classification Head and Loss}
Labels $\hat{y}_{tra_i}$ via linear layer and softmax:
\begin{equation}
	\hat{y}_{tra_i}=\text{Softmax} \left(W_{tra}H_{\text{tra}_i}^{\prime}\right),
\end{equation}
with $W_{tra} \in \mathbb{R}^{2 \times 4d}$. 
Trajectory loss $\mathcal{L}_{\text{tra}}$:
\begin{equation}
\mathcal{L}_{\text{tra}_i}\&=-\sum_j^J y_{tra_i}^j \log \left(\hat{y}_{tra_i}^j\right)\nonumber+\left(1-y_{tra_i}^j\right) \log \left(1-\hat{y}_{tra_i}^j\right),
\label{eq:featureConcatenation1}
\end{equation}

\begin{equation}
\mathcal{L}_{\text {tra}}=\frac{\mathcal{L}_{\text {tra}_1}+\mathcal{L}_{\text {tra}_2}}{2}.
\label{eq:featureConcatenation}
\end{equation}

\subsection{Multi-Task Learning}
We employ multi-task learning, jointly training auxiliary and main tasks. Given the interaction task's greater complexity, we use adaptive weighting~\cite{liebel2018auxiliary}:
\begin{align}
\mathcal{L}&=\frac{1}{2 c_1^2} \mathcal{L}_{\text {inter }} +\frac{1}{2 c_2^2} \mathcal{L}_{\text {tra }}\nonumber\\&+\log \left(1+c_1^2\right)+\log \left(1+c_2^2\right),
\label{eq:mt}
\end{align}
where $c_1$ and $c_2$ are learnable parameters, initialized to 1.

\section{Experiments}
\subsection{Train/Test Split}

We divide our interaction dataset into training, validation and testing with the ratio of 7:1:2. The following section reports various metrics of the test set.

\subsection{Evaluation Metrics}
To quantitatively evaluate our model, we assess its classification performance on the interaction task by computing Accuracy ($\text{Acc}$), Precision ($\text{P}$), Recall ($\text{R}$), and F1-score ($\text{F1}$).

\subsection{Baseline Methods}

In this study, we proposed eight baseline models.

\begin{itemize}    
    \item \textbf{Bi-LSTM~\cite{tao2020character}}: We employ this sequence modeling network to process temporal dependencies in interaction data through bidirectional recurrent layers.
    
    \item \textbf{BERT~\cite{devlin2018bert}}: We include this foundational transformer-based language model as a standard pretraining baseline for comparison.
    
    \item \textbf{R-Bert~\cite{wu2019enriching}}: This BERT extension enhances entity context by explicitly marking target entities and integrating their position-aware representations.
    
    \item \textbf{RoBERTa~\cite{liu2019roberta}}: An optimized BERT variant trained with larger datasets, dynamic masking, and without NSP objective for improved representation learning.
    
    \item \textbf{AoE~\cite{li-li-2024-aoe}}: This state-of-the-art pretrained model outperforms BERT and RoBERTa, achieving top results on MTEB benchmarks for text similarity tasks.
    
    \item \textbf{GPT-4o-mini\footnote{https://platform.openai.com/docs/models\#gpt-4o-mini}}: We use this lightweight LLM with chain-of-thought prompting to evaluate generative reasoning capabilities (prompt details in Appendix).

    \item \textbf{COSMOS~\cite{zhang2025paths}}: COSMOS is used for extracting trajectory triples, but it does not explicitly model entity relationships. To compare with the method proposed in this paper, we retrained it on the interaction quadruple extraction task.

    \item \textbf{COSMOS$_{\text{Frozen}}$~\cite{zhang2025paths}}: Extract interaction quadruple tasks using the COSMOS model trained with trajectory triplet task. If two triples have the same time and position information, they are merged into one interaction quadruple. This is a heuristic method that completely relies on spatio-temporal consistency.
    
\end{itemize}
\subsection{Experimental Results}

We assess the experimental results of our model by comparing it to our introduced baselines on the manually annotated interaction dataset.

\vspace{-0.2em} 
\begin{table}[!htb]
  \centering
    \begin{tabular}{lllll}\toprule
    Methods&Acc (\%) &P (\%)&R (\%)&F1 (\%)\\\midrule
    COSMOS$_{\text{Frozen}}$ &69.32  &52.07 & 72.34 & 60.55\\
    GPT-4o-min & 74.17 & 72.69 & 80.60 & 76.44  \\
    
    Bi-LSTM & 72.06 & 70.67 & 79.10 & 74.65  \\
    
    BERT & 76.61 & 73.33 & 84.43 & 78.49  \\
    
    RoBERTa  & 80.16 & 74.87 & \textbf{89.55} & 81.55   \\
    
    AoE  & 81.60 & 78.49 & 88.65 & 83.28 \\


    COSMOS &81.93  & 81.01 & 85.73   & 83.30\\
    
    R-Bert & 82.37 & 79.69 & 88.70 & 84.01  \\
    
    FALCON & \textbf{85.48} & \textbf{83.67} & \textbf{89.55} & \textbf{86.51}   \\
\bottomrule
    \end{tabular}%
    \vspace{-5pt} 
    \caption{Performance comparison on the test set.}
    \label{tab:baseline_performance1}%
    \vspace{-10pt} 
\end{table}%

\subsubsection{Prediction Performance}

Table~\ref{tab:baseline_performance1} details model performance on the interaction datasets. Our model outperforms all others on every metric, leading the runner-up by 3.11\% accuracy, 2.50\% F1, and 3.98\% precision. It matches RoBERTa for the highest recall.

R-Bert ranks second overall. Although it incorporates entity information, it neglects positional variations within entities. COSMOS (F1=83.30\%) and AoE (F1=83.28\%) follow closely. Despite their innovations (COSMOS combines CNN/BERT with contrastive/semi-supervised learning; AoE uses a complex space loss to avoid vanishing gradients), their limitations confirm the need for explicit entity modeling in this task. RoBERTa (F1=81.55\%) and BERT (F1=78.49\%) perform worse than R-Bert, further underscoring entity information's importance. RoBERTa enhances the model's generalization ability by adopting dynamic masks during training, which might be the reason for its superior recall performance in our task.

In addition, we observe that GPT-4o-mini (F1=76.44\%) surpasses Bi-LSTM (F1=74.65\%), showing promise but still lagging behind specialized supervised methods. COSMOS$_{\text{Frozen}}$ performs worst, failing to capture semantic interactions due to error accumulation.




\textbf{In the Appendix, we also present the implementation details and the generalization study of the models.}

\vspace{-0.2em} 
\begin{table}[!htb]
\setlength\tabcolsep{3pt}
  \centering
    \begin{tabular}{lllll}\toprule
    Methods&Acc (\%) &P (\%)&R (\%)&F1 (\%)\\\midrule
     
    $FALCON_{w/o\ ft \& mt}$ & 83.70 & \textbf{84.80} & 87.42 & 84.80 \\ \\
     
    $FALCON_{w/o \ mt}$ & 83.81 & 80.76 & \textbf{90.40} & 85.31 \\
     
    $FALCON_{w/o \ ft}$ & 84.70 & 82.26 & 89.98  & 85.94 \\
  

 
    
    $FALCON$ & \textbf{85.48} & 83.67  & 89.55  & \textbf{86.51}   \\
\bottomrule
    \end{tabular}%
    \vspace{-5pt}
    \caption{Results of the ablation study.}
    \label{tab:ablation_study}%
    \vspace{-10pt}
\end{table}%

\subsection{Ablation Study}

We conduct an ablation study on the interaction dataset to evaluate the effectiveness of key components, as shown in Table~\ref{tab:ablation_study}. Specifically, FALCON$_{w/o \ ft}$ excludes the feature transfer module, FALCON$_{w/o \ mt}$ omits the multi-task learning strategy, and FALCON$_{w/o \ ft \& mt}$ removes both components. Consistent with our expectations, FALCON achieves the best performance, while FALCON$_{w/o \ ft \& mt}$ exhibits the worst performance, underscoring the effectiveness of each component. 

Additionally, we conduct more detailed ablation experiments on the design of multi-task learning and feature transfer, as shown in the Appendix.

     
     
     
  

 
    

\section{Analysis of Extracted Interactions}
We extract 658,966 spatio-temporal interaction quadruples from the English Wikipedia, selected from a total of 4,337,152 auto-generated candidate interaction quadruples. 

We manually reviewed 300 extracted quadruples and found that 82\% (246 quadruples) were accurate. Additionally, we inspected 100 samples labeled as negative by FALCON, among which 93\% were indeed incorrect, closely aligning with the model's performance on the test set.

From these quadruples, we select 293,586  where each individual has a separate Wikipedia page, providing richer personal attributes. There are 291,136 people and 49,579 locations, and the interactions span from the year 1000 to 2024\footnote{We use the Wikipedia dumps of January 11, 2025 from https://dumps.wikimedia.org/}. 

In the Appendix, we present the complete dataset from the perspectives of geography and spatio-temporal networks. In the following section, we conduct an empirical analysis of intra- and inter-party interactions among US political figures, as a new angle to examine political polarization, thereby demonstrating the potential of our data.

\subsection{Interactions of US Political Figures}
Utilizing the extracted data, we focus on the real-world interactions of US political figures to gain insights into \textit{political polarization}, a significant area of academic interest~\cite{fiorina2008political}. Unlike most large-scale studies that primarily rely on polls ~\cite{fiorina2008political,pew2017partisan,pew2019partisan,holder2023polarizing}, social media sources like Twitter (now X) ~\cite{conover2011political,garimella2017long,flamino2023political,schoenmueller2023frontiers}, or analyzing the behavioral patterns of entrepreneurs with explicit political orientations~\cite{fos2022political}, our analysis offers a unique perspective by examining direct interactions among political actors across time and space.

We identify 14,084 interactions among 3,896 Republicans and 3,995 Democrats from 1960 to 2024. These interactions are categorized as \textit{intra-party}, where individuals belong to the same party, or \textit{inter-party}, where they come from different parties.

We further classify all interactions into three types: \textit{Adversarial} (Conflicting political interests), such as election competition, and debate; \textit{Cooperative} (Joint actions toward common political goals), such as political work cooperation, and face-to-face support; and \textit{Neutral} (Non-political/symbolic interactions), such as personal relationships, and ceremonial meetings. The \textit{Adversarial} and \textit{Cooperative} types align with the divisions found in studies discussing political competition and cooperation activities~\cite{bassan2022party,jost2022cognitive,bendix2017partisan}. However, in the context of Wikipedia biographies, \textit{Neutral} activities encompass additional interactions beyond these categories, as illustrated in the examples above.

Here we apply GPT-4o-mini to perform the classification and it achieves an accuracy of 93.50\% on a manually verified subset of 200 samples. This indicates that, based on the data extracted from our model, LLMs can perform quite well. The resulting statistics are presented in Table~\ref{tab:gpt_type}. Notably, within the same party, there may be \textit{Adversarial} interactions occasionally. We include the prompts in the Appendix as an example for readers who wish to further explore the types of interactions based on our model's results.

\vspace{-0.2em} 
\begin{table}[!htb]
\setlength\tabcolsep{3pt}
  \centering
    \begin{tabular}{lllll}\toprule
        &Intra-party &Inter-party\\\midrule

    \textit{Cooperative} & \:\:\:\:5,182   & \:\:\:\:1,670 \\

    \textit{Adversarial} & \:\:\:\:1,634  & \:\:\:\:2,317   \\

    \textit{Neutral} & \:\:\:\:1,468 & \:\:\:\:1,812   \\

\bottomrule
    \end{tabular}%
    \vspace{-5pt}
    \caption{Statistics on types of interactions among politicians.}
    \label{tab:gpt_type}%
    \vspace{-10pt}
\end{table}%

\subsubsection{Trends for Inter-Party Interactions}
\begin{figure}[htbp]
    \centering
    \includegraphics[width=0.95\linewidth]{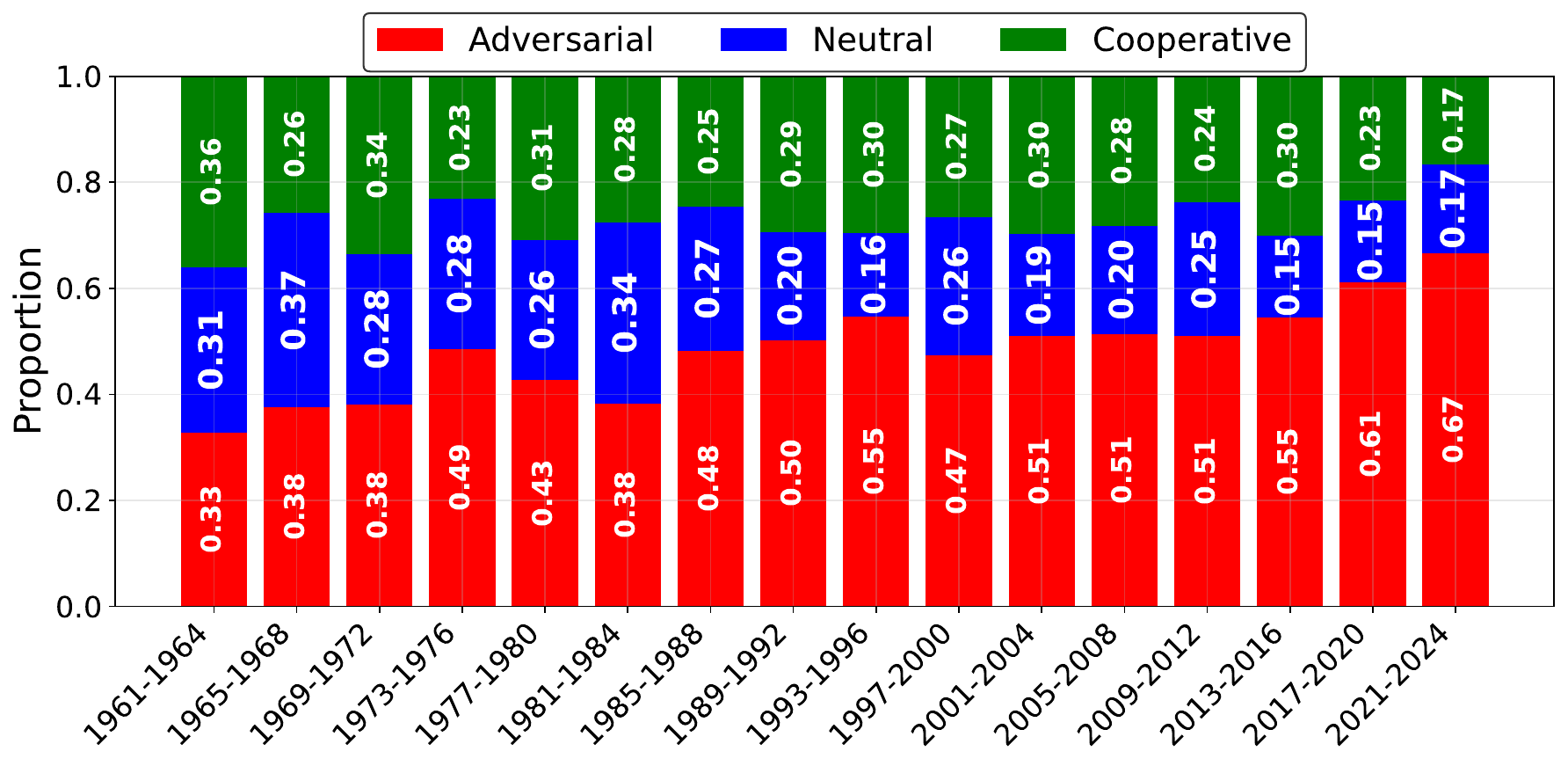}
    \vspace{-5pt}
    \caption{The evolution in the ratios of different types of inter-party interactions.}
    \label{fig:6}
    \vspace{-10pt}

\end{figure}

In our data, the proportion of inter-party interactions (i.e., interactions between different political parties) in total interactions decreased significantly from 0.47 in the 1960s to 0.24 in the 2020s. Furthermore, as shown in Figure~\ref{fig:6}, the ratio of \textit{Adversarial} interactions among all inter-party interactions has been steadily increasing from 1960 to 2024, rising from 32.78\% to 66.67\%. In contrast, the ratios of \textit{Cooperative} and \textit{Neutral} interactions have decreased, falling from 31.09\% to 16.67\% and from 36.13\% to 16.67\%, respectively. This trend likely indicate a growing political polarization over time in the US~\cite{fiorina2008political}.

\vspace{-0.2em} 

\subsubsection{Evolution of Political Interaction Network}
To further capture the complexity of the political interactions, we examine polarization by constructing interaction networks, referencing the classic political polarization study using Twitter~\cite{conover2011political}. 

In our network, each node represents a political figure, and edges between nodes denote interactions. Following the framework of \textit{Weighted Signed Networks} (WSN)~\cite{kumar2016edge}, which assigns signed weights to edges to capture both the nature and strength of interactions, we weight edges as follows: $-2$ for \textit{Adversarial} interactions, $2$ for \textit{Cooperative} interactions, and $1$ for \textit{Neutral} interactions. This weighting scheme reflects our rationale that \textit{Neutral} interactions, though not overtly positive, still signify some degree of goodwill between political figures—hence the non-zero value. Assigning $0$ to \textit{Neutral} interactions would effectively remove them from consideration, which we argue misrepresents their subtle but meaningful role. The values $-2$ and $2$ for \textit{Adversarial} and \textit{Cooperative} interactions, respectively, provide a balanced contrast in polarity and magnitude.

\begin{figure}[htbp]
    \centering
    \includegraphics[width=0.6\linewidth]{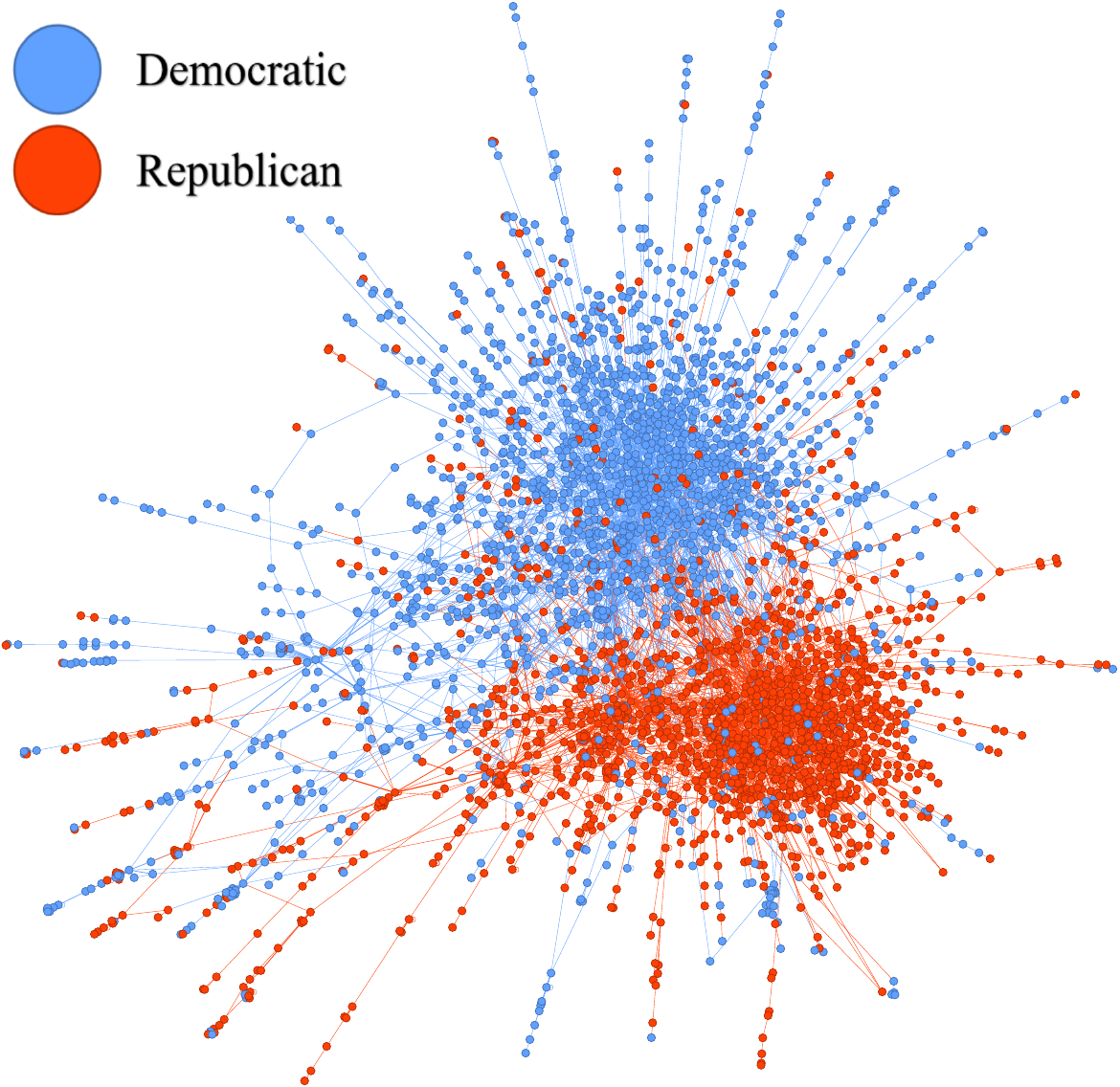}
    \vspace{-5pt}
\caption{US political interaction network (1960-2024) with nodes colored by party (red=Republican, blue=Democrat) and edges weighted by interaction type: \textit{Neutral} (1), \textit{Cooperative} (2), \textit{Adversarial} (-2).}
    \label{fig:7}
    \vspace{-10pt} 
\end{figure}

Figure~\ref{fig:7} visualizes the network of all interactions from 1960 to 2024, where red and blue nodes represent Republican and Democratic figures, respectively. The \textit{ForceAtlas2} layout algorithm is employed to generate an interpretable representation by simulating physical forces, thereby positioning nodes according to their relational dynamics. This approach accentuates the underlying polarization structure within the network.

\vspace{-0.2em} 

To quantify political polarization, we employ \textit{standardized modularity} based on conventional modularity \cite{newman2004finding}, partitioning nodes by party (Republican/Democrat). Higher values indicate stronger party segregation (reduced cross-party interactions), with raw values standardized against randomized networks \cite{conover2011political} to control for size/connectivity effects 
(see Appendix for details), yielding a robust polarization measure where larger values reflect more severe divisions.

\begin{figure}[htbp]
\vspace{-10pt} 
    \centering
    \includegraphics[width=0.9\linewidth]{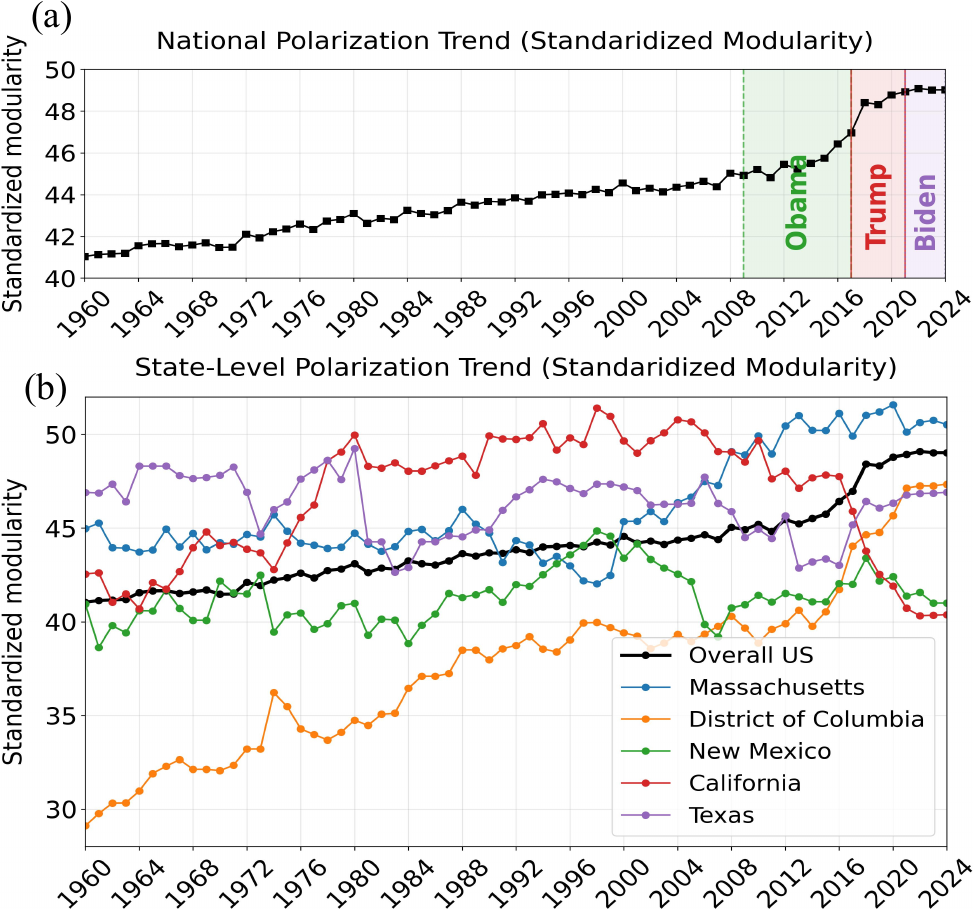}
    \vspace{-5pt} 
    \caption{
(a) National polarization trend (1960-2024) showing acceleration during Obama's second term and peak under Trump; 
(b) State-level trends with D.C. rising fastest (2nd in 2024), Massachusetts leading, and California declining.}
    \label{fig:8}
    \vspace{-20pt} 

\end{figure}

Using annual interaction networks from 1960-2024, we calculate standardized modularity to measure polarization over time (Figure~\ref{fig:8}a). The results reveal a clear upward trend, with accelerated growth during Obama's second term and the sharpest increase occurring under Trump's first term. These findings align with \cite{doherty2017partisan}'s poll-based conclusion that partisan divisions reached record levels during Obama's presidency and expanded further in Trump's initial year.

We analyze state-level networks for the top 5 regions (including Washington D.C. as the nation's political center) using within-state interactions. Figure~\ref{fig:8}b reveals distinct polarization patterns: Massachusetts shows consistent growth to lead in 2024, while Texas and New Mexico remain stable. Most strikingly, D.C. - despite having the lowest initial polarization in 1960 - demonstrates the steepest increase, rising to second highest by 2024. California presents a unique case with significant polarization declines in recent years.

\section{Conclusion}

We propose a new task of extracting spatio-temporal interaction from Wikipedia and introduce FALCON, to effectively extract life spatio-temporal interactions from Wikipedia biography pages by combining the AR-Bert, feature transfer and multitask learning. We also validate its generalizability on Encyclopedia Britannica.
To validate the method and showcase the potential of the extracted data, we analyze US political interactions, offering a new perspective on political polarization.
We hope that the open-sourced code, the extracted interactions, and the \textit{WikiInteraction} ground truth dataset, can support the spatio-temporal interaction extraction research and the analytical studies based on these spatio-temporal interactions. As the largest of its kind, our dataset can be the basis for data-driven grand narratives and explorations of human interaction mechanisms. 

We have to note that since we choose to extract interactions from the English Wikipedia, there can be a bias that the extracted individuals are more likely to be from the English world~\cite{roy2022information}. This should be considered when any research tries to draw conclusions from our dataset. To mitigate this, a possible future step is to extend our framework to versions of Wikipedia in other languages and further explore different designs of extraction algorithms. While our current pipeline uses GPT as a post-processing module for interaction classification, this task-agnostic implementation serves as a proof of concept rather than an optimized solution. Future work will develop dedicated architectures to achieve domain-independent generalization beyond political contexts. Currently, interaction types are generated by GPT and manually verified as a post-processing step. To improve performance in the future, we may develop a dedicated model for this task.


\bibliography{aaai25} 

\clearpage

\section{Appendix}

\subsection{Acquisition of Training Data}

Since there is no available dataset that captures the focus of our study, spatio-temporal interactions, we annotate a new ground truth dataset. The following section details the process of creating the $WikiInteraction$ dataset.

The dataset is derived from the biography sections of the English Wikipedia. We source the list of individuals and the links to their respective Wikipedia biography pages through Wikidata~\cite{moller2022survey}. 

This section details the methodology used to extract and label potential quadruples in the format of (\textit{Person1}, \textit{Person2}, \textit{Time}, \textit{Location}) from the biography pages.

\subsubsection{Extracting Quadruples}

Building upon an existing pipeline which can extract trajectory triplet from text based on the combination of NER and syntex tree~\cite{zhang2025paths}, we construct candidate interaction quadruples $ (\textit{Person1}, \textit{Person2}, \textit{Time}, \textit{Location}) $ by pairing trajectory triplets that co-occurrence (i.e., two triplets share the same time and location).

To evaluate the coverage of our method, we compare the interactions mentioned in the original pages with those extracted from the target text segment. We manually review 12 biography pages containing a total of 103 interaction descriptions. Our extraction pipeline could capture at least 94.00\% of the interactions across these pages. Unidentified interactions are often due to certain interactions require multiple segments of text to infer, making their recognition hard even for humans.

\subsubsection{Annotating}

After we extract candidate quadruples using the above method, we randomly select 4,507 quadruples for manual annotation. 
Furthermore, we also split each interaction quadruple (\textit{Person1}, \textit{Person2}, \textit{Time}, \textit{Location}) into two trajectory triples (\textit{Person}, \textit{Time}, \textit{Location}) for the annotation of trajectory category, since the detection of trajectory is an auxiliary task in our multi-task learning framework. Our annotation has involved three undergraduate students: two annotators and one checker. We annotate 4,507 interaction quadruples and 9,014 trajectory triplets. The distribution of positive and negative samples has been presented in the main text.
\subsection{Implementation Details}
In this paper, we use a BERT-base\footnote{https://github.com/google-research/bert} model to generate word embeddings. While any suitable model can be employed as well, in FALCON , we set $d=768$. We train  FALCON  using the AdamW optimizer with a learning rate configuration of $ 5\mathrm{e}^{-5} $. All the aforementioned experiments are conducted on two RTX 3090 GPUs.

\subsection{Generalization Analysis}
We test how the models trained on Wikipedia generalize to another important source of biographies, Encyclopedia Britannica (EB)\footnote{https://www.britannica.com/Biographies}. We compare our model, FALCON, with the four best performing models from our main experiment: R-Bert, COSMOS, AoE and  RoBERTa, where all models are trained on the original training set constructed from Wikipedia. In Table~\ref{tab:Generalization}, we present the new F1 scores tested on 500 candidate interaction quadruplets from EB, which are manually labeled, alongside the original results tested on Wikipedia from Table~\ref{tab:baseline_performance1}.

All models show a decrease to some extent, while ours achieves an F1 score of 82.76\%, remaining the highest among the models, with a decrease of only 3.75\%. This demonstrates its ability to generalize, surpassing all four baselines.

\begin{table}[!htb]
\setlength\tabcolsep{3pt}
  \centering
    \begin{tabular}{lllll}\toprule
       Methods&F1 (on EB)&F1 (on wiki)&$\Delta$F1\\\midrule
  RoBERTa   & \:\:\:\:74.25&\:\:\:\:81.55 & \:-7.30\\
    AoE   & \:\:\:\:76.83& \:\:\:\:83.28& \:-6.54\\
    COSMOS   & \:\:\:\:78.13& \:\:\:\:83.30& \:-5.17\\
  R-Bert & \:\:\:\:78.48& \:\:\:\:84.01&\:-5.53 \\
 FALCON & \:\:\:\:\textbf{82.76}& \:\:\:\:\textbf{86.51}&\:\textbf{-3.75}  \\ 
\bottomrule
    \end{tabular}%
    \caption{Models trained on Wikipedia and tested on Encyclopedia Britannica and Wikipedia, respectively.}
    \label{tab:Generalization}%
\end{table}%

\subsection{Detailed Ablation}

\begin{table}[!htb]
\setlength\tabcolsep{3pt}
  \centering
    \begin{tabular}{lllll}\toprule
    Methods&Acc (\%) &P (\%)&R (\%)&F1 (\%)\\\midrule

    $FALCON_{w/o \ aw}$ & 84.26   & 81.20  & 89.34   & 85.51   \\


    $FALCON_{concat}$ &85.25  &\textbf{84.85}  & 87.21 &86.01    \\


 
    
    $FALCON$ & \textbf{85.48} & 83.67   &\textbf{89.55}  & \textbf{86.51}   \\
\bottomrule
    \end{tabular}%
    \caption{Specific ablation study.}
    \label{tab:Specific_ablation_study}%
\end{table}%

We conduct specific ablation studies on multi-task learning and feature transfer, as shown in Table~\ref{tab:Specific_ablation_study}. For multi-task learning, we validate the effectiveness of the adaptive task weighting strategy. Specifically, FALCON$_{\text{w/o aw}}$ denotes the model without the adaptive task weighting strategy. The results show a significant decrease in overall metrics. The adaptive weighting strategy enables the model to automatically focus on more challenging interaction tasks, thereby enhancing its performance on this task. Regarding feature transfer, we assess the effectiveness of the feature fusion module. FALCON$_{\text{concat}}$ represents the model where the feature fusion module is removed and trajectory features and interaction features are merely concatenated. 
We observe that compared to FALCON$_{\text{w/o ft}}$ (refer to Table~\ref{tab:ablation_study}), which does not utilize any feature transfer strategy, FALCON$_{\text{concat}}$ only achieves a marginal improvement in the F1 score. Although precision increases by 2.59\%, recall decreases by 2.77\%. This indicates that the concatenation strategy in our task trades recall for precision. In contrast, FALCON, which employs a feature fusion module in its feature transfer strategy, manages to robustly increase precision while only slightly sacrificing recall, thereby achieving better improvements in the overall metrics of F1 and recall.

\subsection{Prompt for GPT-4o-mini Baseline}
\begin{figure}[h]
    \centering
    \includegraphics[width=1 \linewidth]{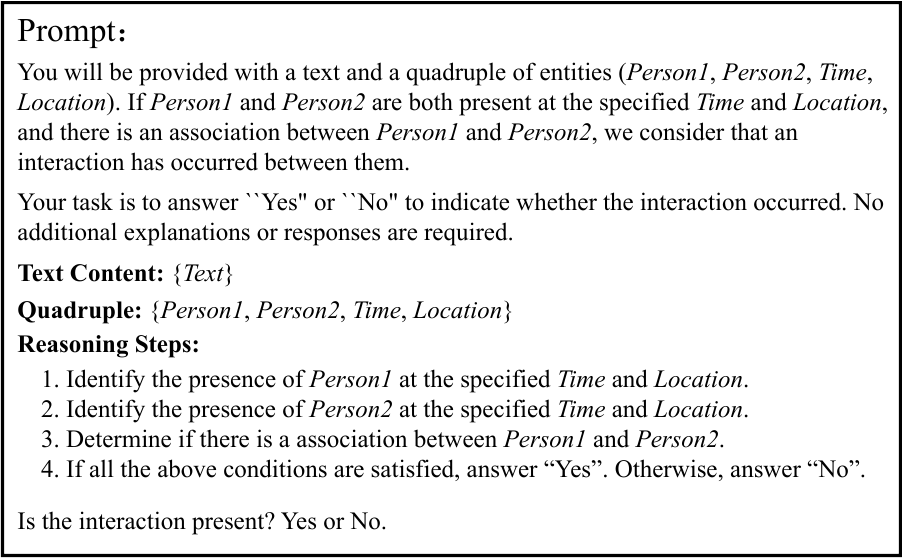}
    \caption{The prompt for GPT-4o-mini Baseline.}
    \label{fig:c_p}
\end{figure}

Figure~\ref{fig:c_p} shows the prompt provided when we use GPT-4o-mini (gpt-4o-mini-2024-07-18). The temperature here is set to 1 and one trial is performed for each input.

\subsection{Prompt for Political Interaction Classification}
\begin{figure}[h]
    \centering
    \includegraphics[width=1 \linewidth]{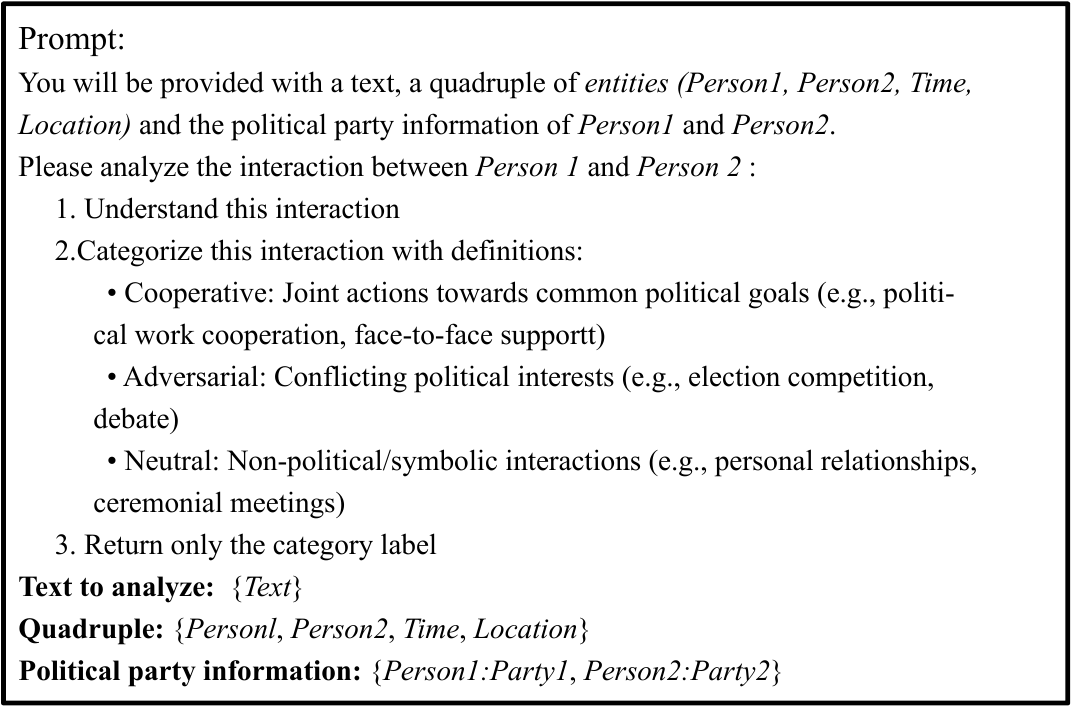}
    \caption{The prompt for GPT-4o-mini to categorize political interactions.}
    \label{fig:c_p2}
\end{figure}

We employ GPT-4o-mini to categorize political interactions into three distinct classes: \textit{Adversarial}, \textit{Cooperative}, and \textit{Neutral}, maintaining identical experimental conditions to the baseline configuration. The prompt is provided in Figure~\ref{fig:c_p2}.

\subsection{Calculation of Standardized Modularity}

Following Conover et al. \cite{conover2011political}, we implement their normalized modularity calculation method. As direct comparison of modularity values across networks with varying sizes and connection densities is challenging, their approach evaluates the relative quality of cluster assignments. The implementation involves:
\begin{itemize}
    \item Generating $N=1000$ randomized network samples preserving both degree sequence and edge weight distribution (for weighted graphs in our scenario)
    \item Computing the mean ($\mu$) and standard deviation ($\sigma$) of modularity $Q$ from randomized samples
    \item Deriving the Z-score: $Z = (Q_{\text{original}} - \mu) / \sigma$
\end{itemize}
This normalization procedure quantifies how significantly the observed community structure deviates from random expectation, while maintaining the original network's topological characteristics through degree- and weight-preserving randomization.

\subsection{Additional Analysis}

\subsubsection{Overall Analysis}
\begin{figure}[htbp]

    \centering
    \includegraphics[width=1\linewidth]{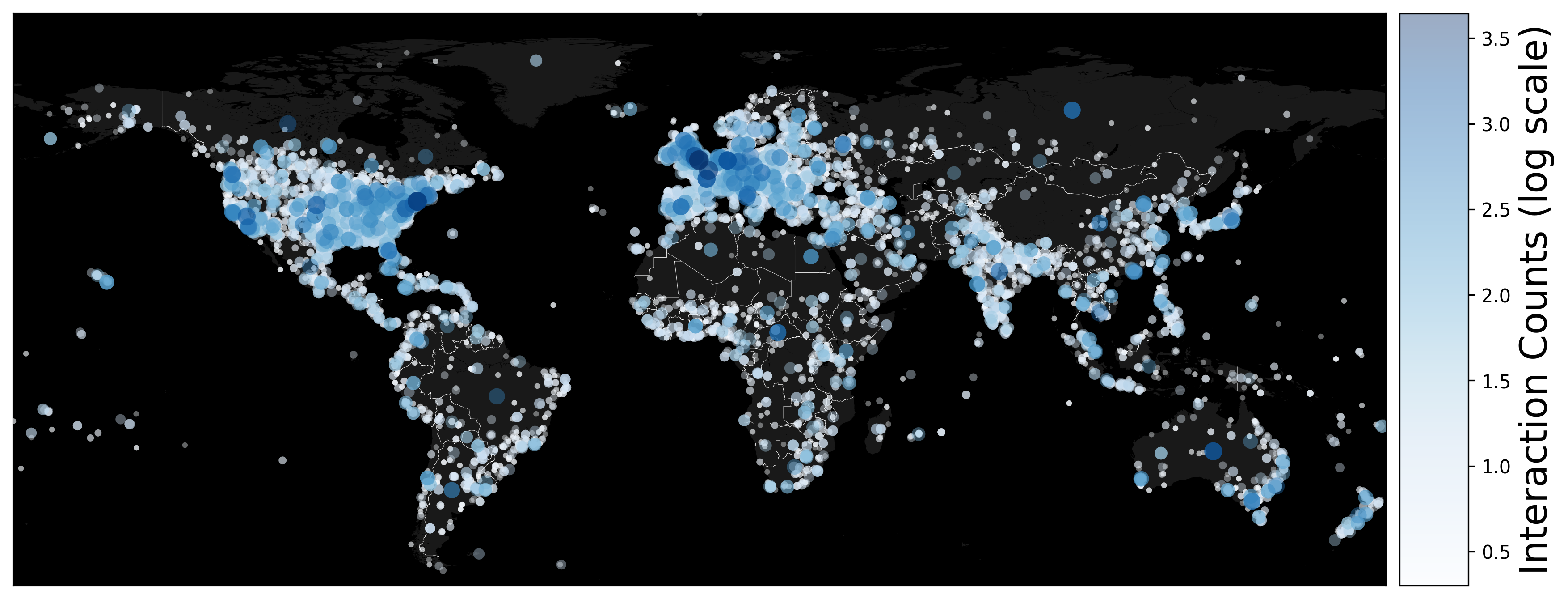}
    \caption{A geographic heatmap of interaction locations, where darker colors indicate a higher number of interaction events.}
    \label{fig:geographic_heatmap}
\end{figure}

Figure~\ref{fig:geographic_heatmap} illustrates the global distributions of these interactions, showing that most interactions occur in North America and Europe, followed by Southeast Asia and Australia.


\begin{figure}[!t]
    \centering
    \includegraphics[width=1\linewidth]{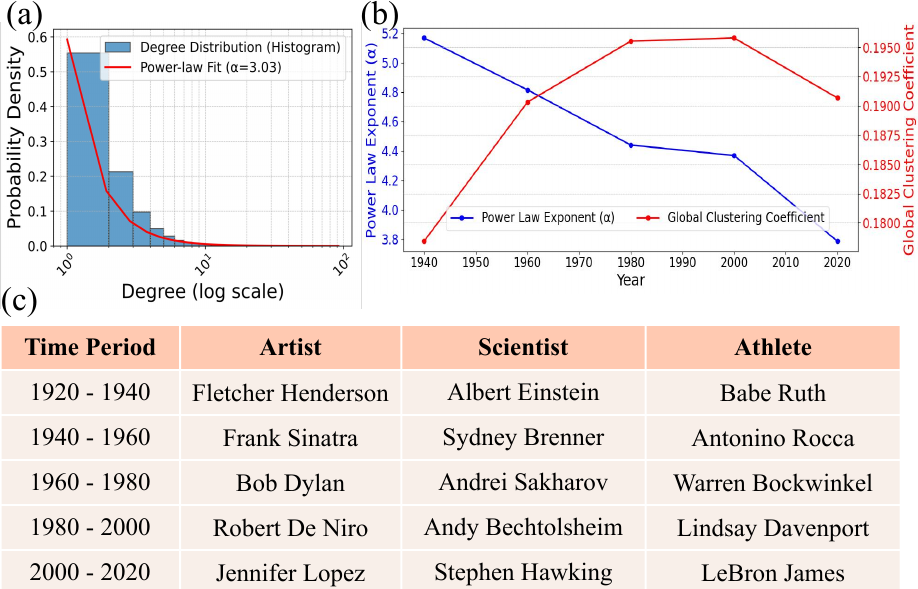}
    \caption{(a) In the constructed network, the distribution of node degrees adheres to a power-law distribution. (b) Clustering coefficient \( c \) and power-law distribution parameter \( \alpha \) over 20-year intervals from 1920 to 2020. (c) From 1920 to 2020, every 20 years, the top-ranked individuals in the professions of artists, scientists, and athletes according to their PageRank.}
    \label{fig:time}

\end{figure}
With these interactions, we then build an interaction network. In Figure~\ref{fig:time} (a), we illustrate the degree distribution across the entire network, observing a conforming pattern to the power-law distribution. 

We further examine the temporal perspective. Using data from each 20-year interval from 1920 to 2020, we construct 5 sub-networks. For each sub-network, we calculate the clustering coefficient $c$ and the power-law distribution parameter $\alpha$. As we can see from Figure~\ref{fig:time} (b), $c$ exhibits an increasing trend, while the $\alpha$ shows a decreasing trend, indicating the interaction networks are becoming centralized over time.

For each 20-year interval, we identify the top individuals in the occupations of artist, scientist, and athlete, respectively, ranked by PageRank scores. This is illustrated in Figure~\ref{fig:time} (c).

\begin{figure}[!t]
    \centering
    \includegraphics[width=1.02\linewidth]{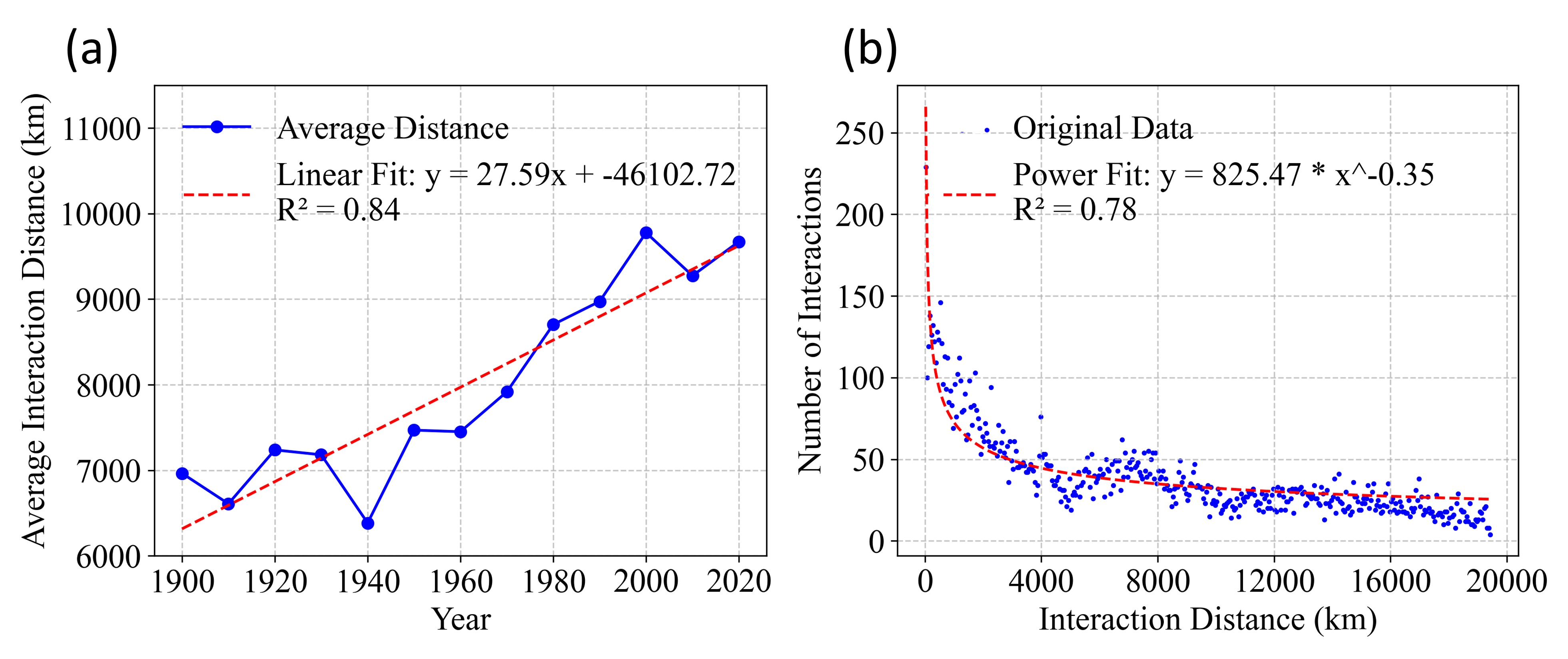}
    \caption{(a) From 1900 to 2020, the evolution of the average interaction distance. (b) The distributional relationship between the number of interactions and the distance of these interactions. }
    \label{fig:distance}
\end{figure}

\subsubsection{Interaction Distance}
We define interaction distance as the cumulative distance from \textit{Location} to both \textit{Person1}'s and \textit{Person2}'s birthplaces, indicating the travel distance required for the interaction. As illustrated in Figure~\ref{fig:distance} (a), the average interaction distance exhibits a growing trend over time, likely attributable to advancements in modern transportation and communication. In contrast, Figure~\ref{fig:distance} (b) shows that most interactions occur over short distances, with the distribution conforming to a power-law pattern. These findings are consistent with those reported by \citet{illenberger2013role}.

\begin{figure}[!t]
    \centering
    \includegraphics[width=1\linewidth]{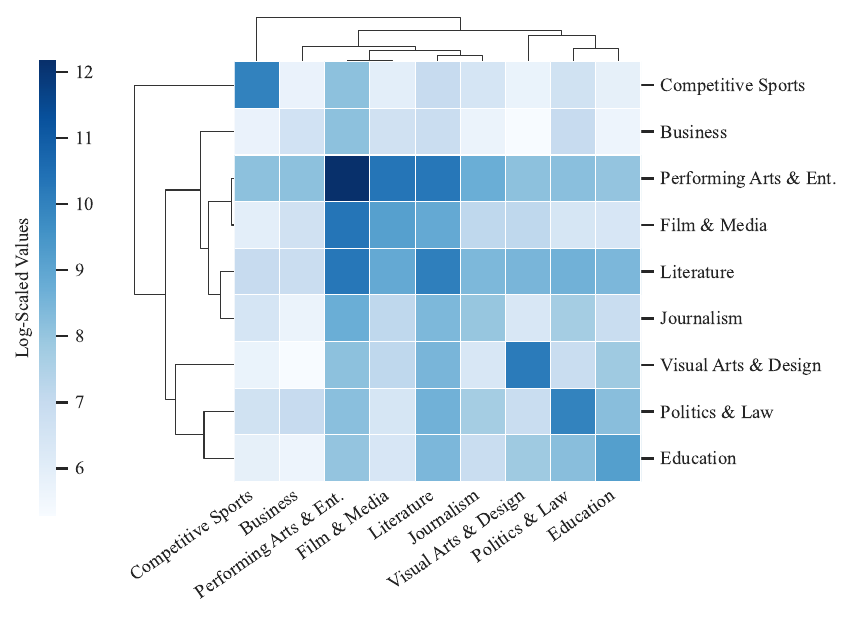}
    \caption{Interactions among different professions are illustrated. The diagonal of the matrix reflects the internal interactions within each individual occupation. The rows and columns represent the professions engaging in the interaction, with the color intensity indicating the volume of interactions. The hierarchical clustering groups related professions, indicating similar interaction patterns.}
    \label{fig:c_1}
\end{figure}

\begin{figure}[!t]
    \centering
    \includegraphics[width=1.05\linewidth]{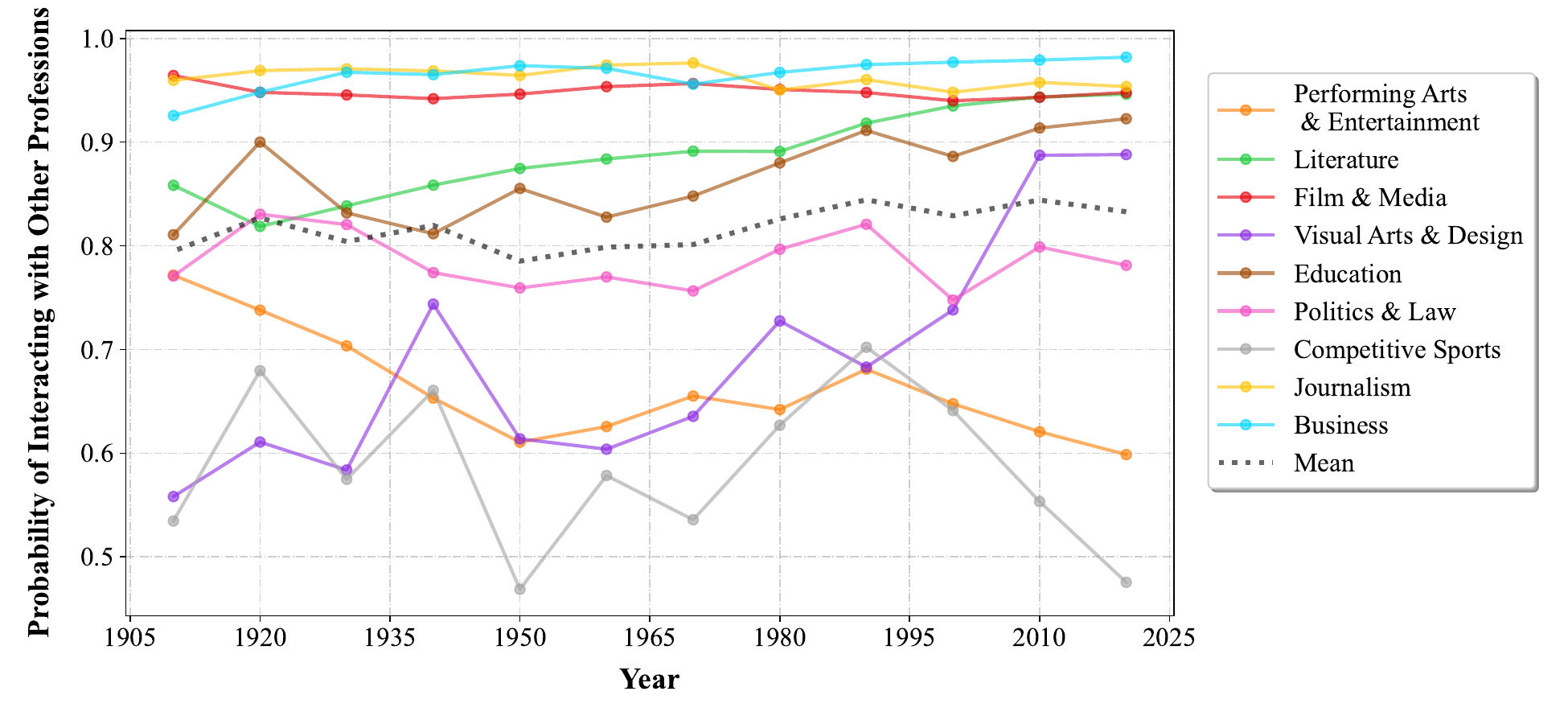}
    \caption{The evolution of interaction probabilities between each profession and the rest over time, along with the depiction of average interaction rates.}
    \label{fig:c_2}
\end{figure}

\subsubsection{Inter-Profession Interactions}
We examine how individuals from different professions interact. We categorize the individuals into 9 professions, and the frequency of interactions between these professions is illustrated in Figures~\ref{fig:c_1}. The pairs with high interaction frequency include ``Politics \& Law'' vs. ``Education'', ``Journalism'' vs. ``Literature'', and ``Performing Arts \& Ent'' vs. ``Film \& Media'', while the ``Competitive Sports'' and ``Education'' appear to have minimal interactions.

From Figures~\ref{fig:c_1}, we also observe that certain professions are more likely to interact with others, such as ``Business'', rather than engaging primarily with individuals from their own fields. We quantify this tendency by calculating the ratio of interactions that occur outside of a given profession. We track how these ratios change over time for the 9 professions, as illustrated in Figure~\ref{fig:c_2}. In addition to ``Business'', ``Film \& Media" and ``Journalism" also demonstrate a greater propensity for external interactions. Notably, the visual arts profession shows a growing trend over the years, which may be associated with the rise in interdisciplinary collaborations.



\end{document}